\documentclass[final,5p,times,twocolumn,compress]{elsarticle}

\usepackage{amssymb}
\usepackage{amsmath}
\usepackage{graphicx}
\usepackage{float}
\usepackage{dcolumn}
\usepackage{multirow}
\usepackage{changes}
\usepackage[utf8]{inputenc}
\usepackage[T1]{fontenc}
\usepackage[colorlinks = true,linkcolor = blue,urlcolor  = blue,citecolor = blue,anchorcolor = blue]{hyperref}
\usepackage{enumerate}
\usepackage{times}
\usepackage{microtype}
\usepackage{orcidlink}
\usepackage[percent]{overpic}
\usepackage{stfloats}

\usepackage{ulem} % For strike-through text

\journal{Physics Letters B}

\begin{document}
\begin{frontmatter}

\title{ Density dependent speed of sound and its consequences in neutron stars.
}

\author[1,2]{Suman Pal,\orcidlink{0009-0000-5944-4261}}
\ead{sumanvecc@gmail.com}
\author[1,2]{Gargi Chaudhuri,\orcidlink{0000-0002-8913-0658}}
\ead{gargi@vecc.gov.in}
\address[1]{Physics Group, Variable Energy Cyclotron Centre, 1/AF Bidhan Nagar, Kolkata 700064, India}

\address[2]{Homi Bhabha National Institute, Training School Complex, Anushakti Nagar, Mumbai 400085, India}

\begin{abstract}
We  introduce a  parametrized density-dependent speed of sound and  construct an ensemble of equations of state for neutron stars which are found to closely resemble the realistic equations of state calculated using relativistic mean field theory.  We show that each of these parameters display an unique feature relevant to the properties of the compact stars.  The emergence of special points in the Mass-Radius plot is a significant outcome for neutron stars which is more commonly seen in case of hybrid stars.    We have  also shown that the curvature term in the speed of sound changes its sign for these hadronic equations of state without the matter  reaching the conformal limit or undergoing any phase transition. It is related to the 1st derivative of the energy per nucleon reaching a maximum.  We have also examined the detailed behavior of the trace anomaly and polytropic index for  RMF models, as well as for a density-dependent parametrized speed of sound. Our analysis demonstrates that the sign of the trace anomaly at high densities is sensitive to the stiffness or softness of the EOS.  Different observational constraints from mass-radius and tidal deformability can restrict  the range of parameters in the proposed speed of sound model.

\end{abstract}

%\begin{keyword}
%% keywords here, in the form: keyword \sep keyword, up to a maximum of 6 keywords
%Neutron stars \sep Equation of the state \sep Speed of sound   

%\end{keyword}

\end{frontmatter}

\section{Introduction}
Compact stars act as  remarkable astrophysical laboratories for investigating dense nuclear matter. The comprehensive study of pulsars \cite{Fonseca:2021wxt,Riley:2019yda,Miller:2019cac,Miller:2021qha}, along with the detection of gravitational waves \cite{LIGOScientific:2018cki}, has already provided valuable insights and constraints on the nuclear equation of state (EOS).
The speed of sound ($C_s^2$) \cite{Tan:2021nat,Greif:2018njt,Tews:2018kmu,Fujimoto:2022ohj,Annala:2023cwx,Marczenko:2022jhl,Brandes:2022nxa,PhysRevC.99.035803,Dutra:2015hxa,Zacchi:2015oma,Motornenko:2019arp,Ecker:2022xxj,Sorensen:2021zme,Kojo:2021hqh,Huang:2022mqp,Chiba:2023ftg,Chiba:2024cny,Marczenko:2023txe}, is an important quantity intrinsic to all thermodynamic systems. In the domain of dense nuclear matter the speed of sound holds particular significance for neutron star research. 
Another important quantity is the normalized trace anomaly  ($\Delta$) \cite{Fujimoto:2022ohj,Annala:2023cwx}. Recently, Fujimoto et al.\cite{Fujimoto:2022ohj} explored the trace anomaly as a signature of conformality in neutron stars.

 It can be shown using thermodynamic identities\cite{Marczenko:2023txe}that the speed of sound can be characterized by the slope ($\alpha$) and curvature ($\beta$) of the energy per particle. Recent studies \cite{Marczenko:2023txe} have highlighted that a change in the sign of $\beta$ signals the onset of strongly coupled conformal matter in the cores of neutron stars and hence might be attributed to the change in medium composition at higher densities. 

In this Letter, in order to explore the above fact, we consider the relativistic mean field models and also we  have introduced an  energy-dependent speed of sound to describe the hadronic equation of state, being motivated by the constant speed of sound parameterization \cite{Alford:2013aca,Alford:2017qgh,Li:2021sxb} often used in the description of quark matter. The speed of sound has been parameterized in a manner that closely replicates the behavior observed in well-established relativistic mean field models (RMF) of hadronic matter.
  RMF models \cite{glendenning2012compact,Dutra:2015hxa,nl3,gm1,iufsu,Tolos:2016hhl,dd2,ddme2} can describe a wide range of neutron star equations of state, incorporating components like neutrons, protons, electrons, muons, hyperons, or delta baryons, the stiffness  of the equation of state depending on the choice of RMF model coupling constants.
Although several sound speed models exist in the literature \cite{Greif:2018njt,Tews:2018kmu}, this model offers the advantage of closely matching RMF models while being easy to implement. Its parameterized form allows for convenient tuning of the equation of state by adjusting model parameters.
  
The most interesting part of our parametrization being that each of the three parameters has been observed to be  uniquely connected to some properties of the compact stars. The polytropic index $\gamma$ has been observed to be independent of one parameter while the compactness corresponding to the maximum mass is independent of another parameter. The change in sign of the thermodynamic quantities $\beta$ and $\Delta$ depends on the stiffness of the EoS. Appearance of special points \cite{yudin2014special,Sen:2022lig,Pal:2023quk} by varying the third parameter in the Mass-Radius plot  is an important feature not much discussed before in the literature in the context of hadron stars (without  phase transition). While it is true that  mass-radius relation can be measured observationally, the concept of a special point remains valuable from a theoretical and phenomenological perspective.
Apart from the $C_s^2$ parametrization, we also examine other key quantities related to the equation of state, such as the normalized trace anomaly ($\Delta$) and its logarithmic rate of change ($\Delta'$) with respect to the energy density.

In this letter, we have shown that the sign change in 
 $\beta$ (curvature term)does not necessarily provide a definitive signature of a phase transition or confirm its role as an order parameter.  It can change sign even without the matter reaching the conformal limit or without undergoing any phase transition in terms of the composition. The change in sign in $\beta$ is related to the slope (1st derivative) of energy per particle reaching its maximum or in other words, the velocity of sound being equal to the term $\alpha$ (slope term). We have shown that this can happen for the hadron equations of state for a wide range of parameters. We have also demonstrated the behavior of the trace anomaly and shown that its sign change occurs when $ P > 3\varepsilon $, not necessarily when $ C_s^2 = 1/3 $.

\begin{figure*}[htp] 
    \centering
     \includegraphics[width=0.95\textwidth]{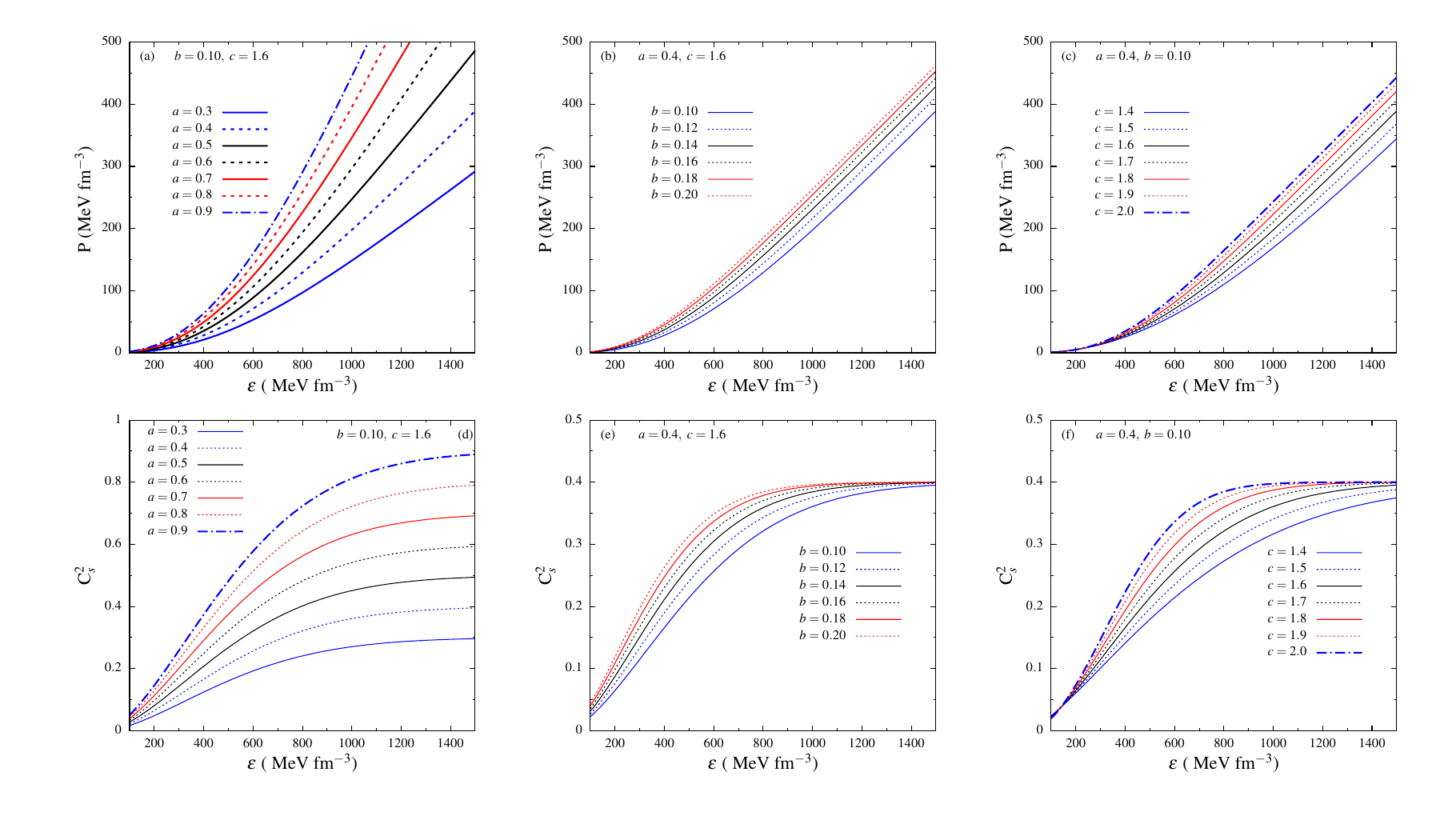}
    \caption{ The upper panel shows the variation of pressure with energy density, while the lower panel shows the variation of $C_s^2$ with energy density.}   
    \label{fig:eos_params}
\end{figure*}  

\begin{figure}[htbp] 
	\centering
	\includegraphics[width=0.45\textwidth]{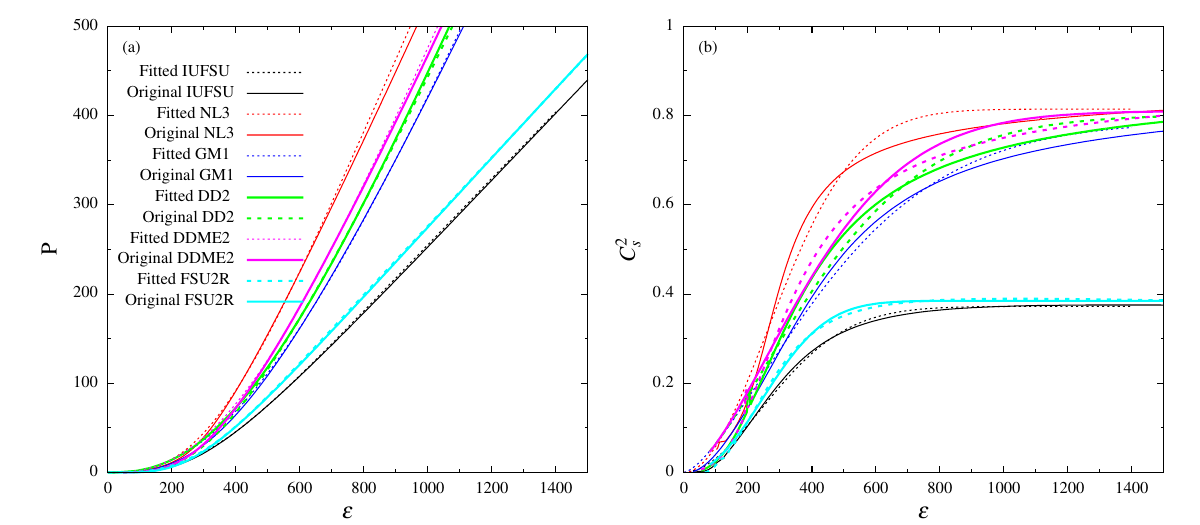}
	\caption{Variation of pressure (a) and $C_s^2$ (b) with energy density for the original and fitted EOS}   
	\label{fig:fit_cs2}
\end{figure}

\begin{figure}[htbp] 
	\centering
	\includegraphics[width=0.45\textwidth]{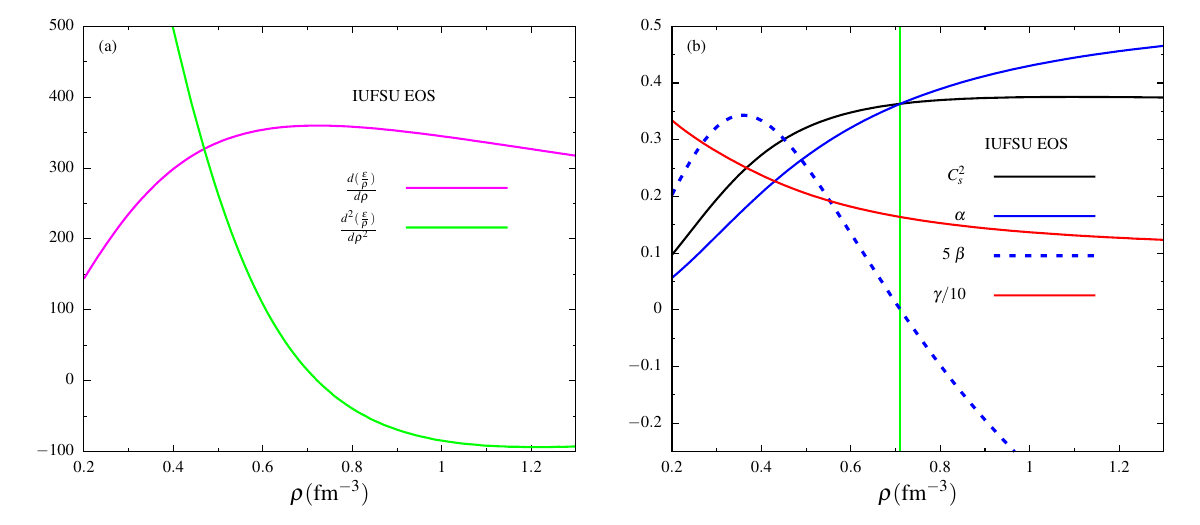}
	\caption{(a) Variation of slope energy density and variation of curvature energy density with density, and (b) variation of $C_s^2$, $\alpha$, $5\beta$ and $\gamma/10$ with density and the green vertical line passes through $\alpha$ equals to $C_s^2$ and $\beta=0$}   
	\label{fig:iufsu_curv}
\end{figure}

\section{Formalism}
The ensemble of equations of state for hadronic matter within a neutron star is characterized by a parameterized speed of sound, defined by three parameters $a$, $b$ and $c$.
\begin{equation} \label{eq:cs2params}
    C_s^2=a\left[1-exp\left(-b \left(\frac{\varepsilon}{\varepsilon_0}\right)^{c}\right)\right]
\end{equation}
where $\varepsilon_0$ is the saturation energy density (in this work we have taken $\varepsilon_0=$140 $\text{MeV fm}^{-3}$). We begin by organizing the set of equations of state for neutron star matter using the speed of sound parametrization outlined above. The equation of state is then obtained through a straightforward integration given by :
\begin{equation}\label{eq:def_cs2_eos}
\text{Speed of sound:~}C_S^2=\frac{dP}{d\varepsilon};~~\text{EOS:}~P(\varepsilon)=\int_0^{\varepsilon}  C_s^2(\bar{\varepsilon}) d \bar{\varepsilon}
\end{equation}
In Fig.~\ref{fig:eos_params}, in the upper panel we have plotted the equation of state and in the lower panel we plotted the speed of sound varying one of the three parameters at a time, keeping the other two fixed. 
The parameter $a$  controls the  stiffness as well as the saturation value of the speed of sound at high densities as is seen from Fig.~\ref{fig:eos_params}.  This puts a restriction on the maximum value of $a$  ( $<$ 1) from the causality limit \cite{glendenning2012compact}. The other two parameters $b$ and $c$  dictates the  pattern of rise of $C_s^2$ from minimum to their saturation values  as is seen from the middle and 3rd column. For lower values of both $b$ and $c$, $C_s^2$ reaches its saturation values at higher energy densities and as these two parameter values are increased, the speed of sound saturates at relatively lower densities as is seen from the figures. The parameter $a$ is varied from 0.3 to 0.9, $b$ from 0.10 to 0.20 while $c$ is being varied from 1.4 to 2.0. 

\begin{table*}[t!] 
\centering
\caption{Fitting parameters of RMF EOS, maximum mass, and corresponding radius.}
\setlength{\tabcolsep}{10pt}
\begin{tabular}{cccccc}
\hline
\hline
EOS&$a$&$b$ & $c$ &$M_{max}$&$M_{max}^{\text{fit}}$  \\
\hline
NL3\cite{nl3}&0.8145&0.1461&1.9579&2.7262&2.7611\\
GM1\cite{gm1}&0.7865&0.1412&1.4617&2.3844&2.4298\\
IUFSU\cite{iufsu}&0.3723&0.1700&1.9086&1.9013&1.9171\\
FSU2R \cite{Tolos:2016hhl} & 0.3847 & 0.1611 & 2.2226&2.0160&2.0335\\
DD2\cite{dd2}&0.8016&0.1411&1.5342&2.4359&2.4955\\
DDME2\cite{ddme2}&0.8099&0.1430&1.6174&2.5074&2.5596\\
\hline
\end{tabular}
\label{tab:1}
\end{table*}
We can decompose the speed of sound as the sum of two terms \cite{Marczenko:2023txe} by using the thermodynamic identities
: $d\varepsilon=\mu d\rho$ and $P=\rho^2\frac{d}{d\rho}\left(\frac{\varepsilon}{\rho}\right)$ : 
\begin{equation} \label{eq:cs2_decompose}
    C_s^2=\frac{1}{\mu}\left(\frac{dP}{d\rho}\right)=2 \frac{\rho}{\mu}\frac{d}{d \rho}\left(\frac{\varepsilon}{\rho}\right)+\frac{\rho^2}{\mu}\frac{d^2}{d \rho^2}\left(\frac{\varepsilon}{\rho}\right)
\end{equation}
the 1st term  on the right hand side  proportional to the first derivative (slope) of energy density and the 2nd being proportional to the 2nd derivative (curvature) of energy density . One can rewrite this as follows. 
\begin{equation}
    C_s^2=\alpha+\beta 
\end{equation} 
Another important quantity that relates the energy density and pressure in a star is the polytropic index ($\gamma$) \cite{glendenning2012compact,Marczenko:2023txe}. It can be expressed as the logarithmic derivative of pressure and energy density as follows:
\begin{equation} \label{eq:poly_index}
    \gamma=\frac{d~log~P}{d~log~\varepsilon}=\frac{\varepsilon}{P}C_s^2 
\end{equation} 
We can relate the parameters $\alpha$ and $\beta$ to the speed of sound and polytropic index as indicated below. 
\begin{equation}
    \alpha=\frac{2 C_s^2}{C_s^2+\gamma},~~\beta=C_s^2-\alpha
\end{equation} 
The curvature term  changes its sign after becoming zero at a certain value of energy density, the slope(1st derivative) of the energy particle reaching its maximum there. When this happens ( $\beta=0$ ) then it can be shown that the following relations hold.
\begin{equation}
    \alpha=\frac{2P}{P+\varepsilon},~~~\gamma=\frac{2\varepsilon}{P+\varepsilon}~~\text{and}~\frac{dP}{d\rho}=\frac{2P}{\rho}
\end{equation}

The last equation shows that the 1st derivative of pressure with density is nonzero. $\beta$ becomes negative when the value of $\alpha$ surpasses the speed of sound. As will be shown in the results section, this can happen irrespective of any peak in the speed of sound or without the matter reaching the conformal limit. Using realistic equations of state as well as the parametrized ones it has been observed that this can occur at few times the nuclear saturation density without any phase transition whatsoever.

The trace anomaly, scaled by the energy density, has recently garnered significant attention in the study of neutron stars. A particularly useful quantity in this context is \(\Delta'\), defined as:  

\begin{equation}
	\Delta=\frac{1}{3}-\frac{P}{\varepsilon},~~~\Delta^{\prime}=\frac{1}{3}-\Delta-C_s^2~~.
\end{equation}  

By combining \(\Delta\) and \(\Delta^{\prime}\), one can construct a single characteristic quantity:  

\begin{equation}
	d_c=\sqrt{\Delta^2+{\Delta^{\prime}}^2}.
\end{equation}

This parameter provides a comprehensive measure incorporating both the deviation from conformal behavior and the influence of the speed of sound.
\begin{figure*}[htp] 
	\centering
    \includegraphics[width=0.30\textwidth]{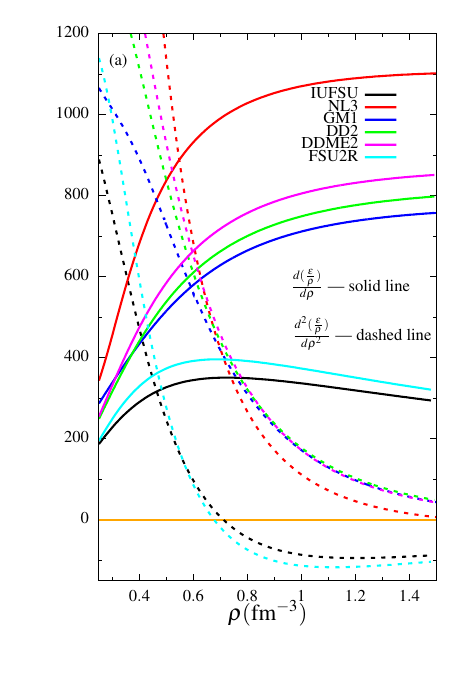}
    \includegraphics[width=0.30\textwidth]{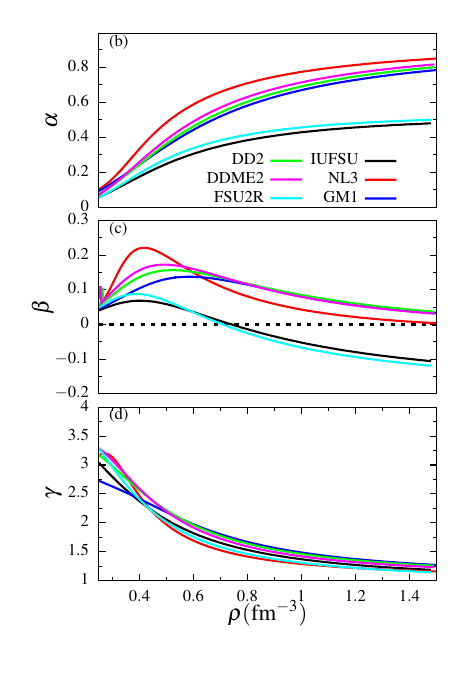}
	\includegraphics[width=0.30\textwidth]{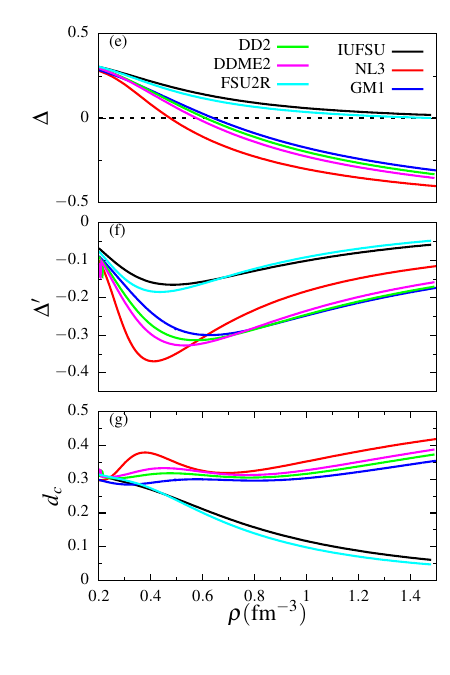}
	\caption{Variation of (a) slope energy density and  curvature energy
density (b) $\Delta$, (c) $\Delta^{\prime}$, and (d) $d_c$ with density for the six selected RMF EOS models.}   
	\label{fig:trace_rmf}
\end{figure*}

\begin{figure*}[htbp] 
    \centering
    \includegraphics[width=0.45\textwidth]{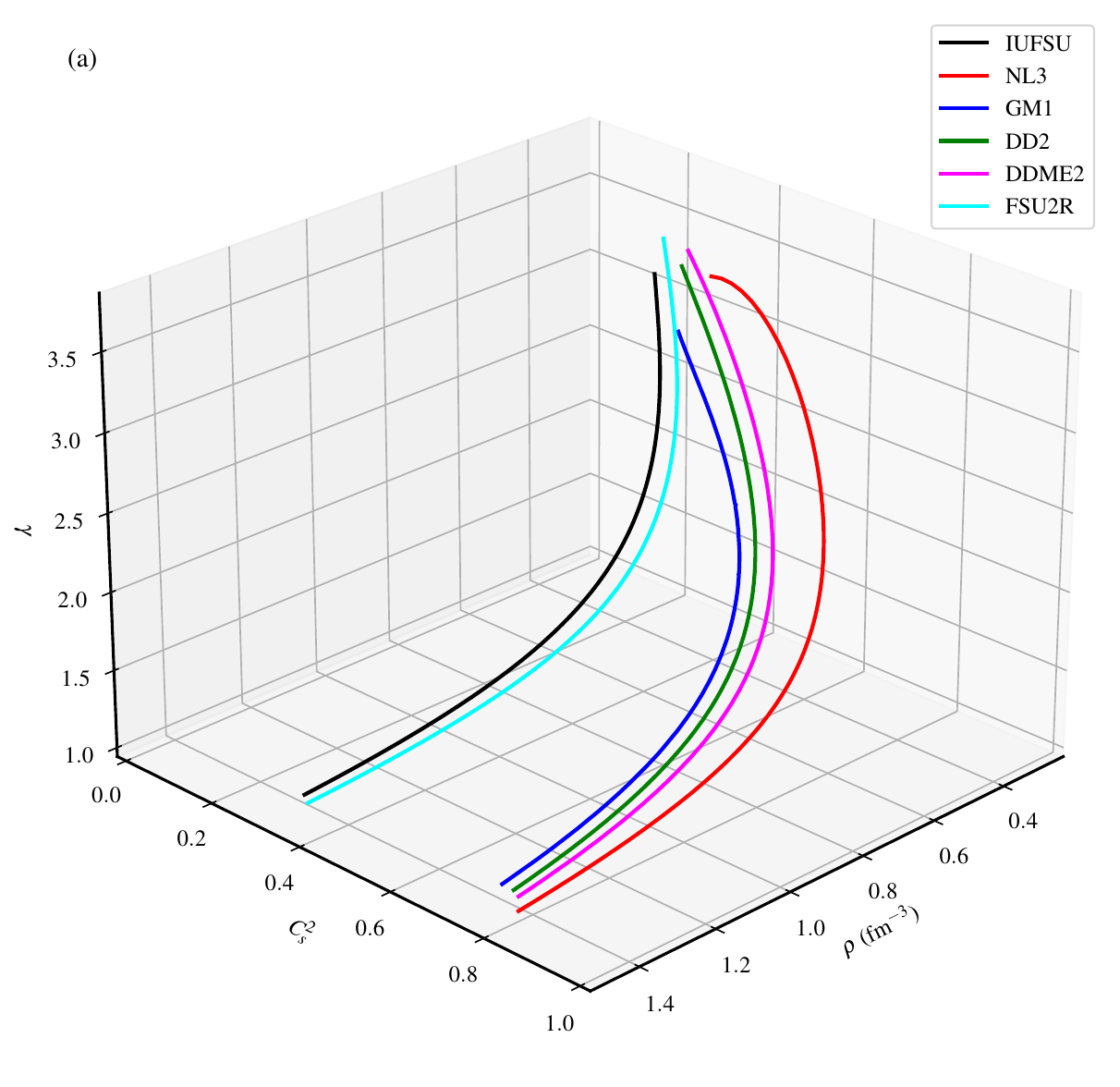} 
    \includegraphics[width=0.45\textwidth]{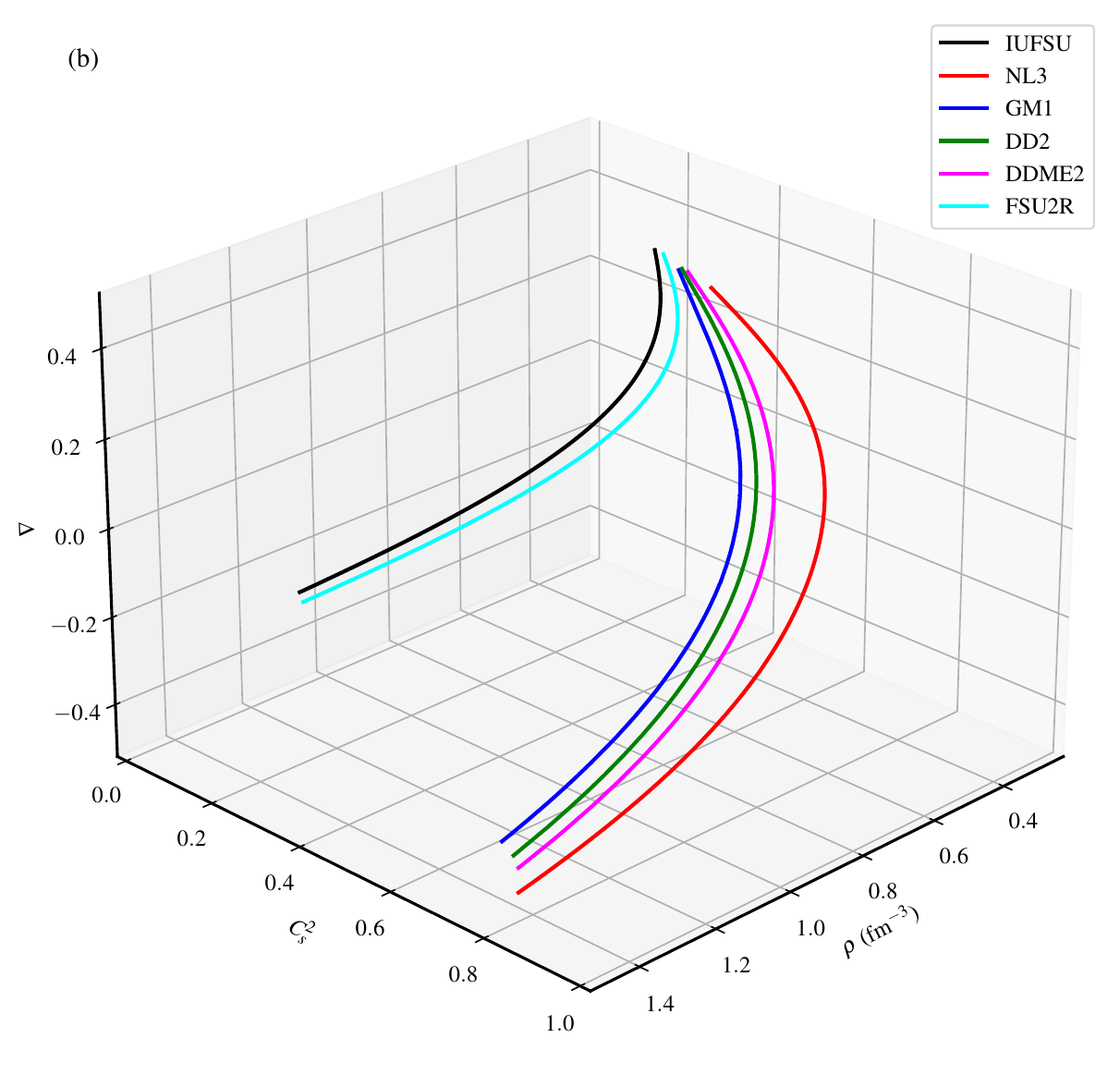}
    \caption{A 3D plot showing the variation of (a) $\rho$, $C_s^2$, and $\gamma$, and (b) $\rho$, $C_s^2$, and $\Delta$.}   
    \label{fig:3d_gamma_delta}
\end{figure*}

Using these EOS, we have calculated the structural properties of neutron stars under static conditions, including the gravitational mass ($M$) and radius ($R$), by solving the Tolman-Oppenheimer-Volkoff (TOV) equations \cite{Tolman_39,Oppenheimer:1939ne}, which describe the hydrostatic balance between gravitational forces and the internal pressure of the star. The dimensionless tidal deformability ($\Lambda$) is determined based on the mass, radius, and tidal Love number ($k_2$) \cite{Hinderer:2007mb,Hinderer:2009ca}.

\begin{figure*}[htbp] 
    \centering
    \includegraphics[width=0.90\textwidth]{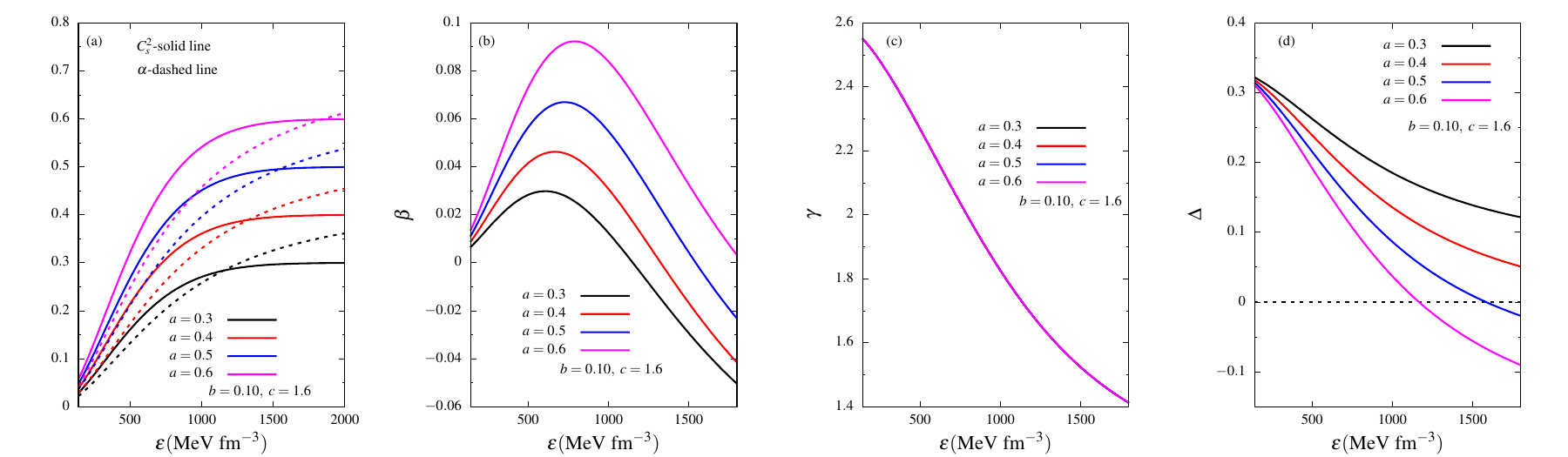} 
    \includegraphics[width=0.90\textwidth]{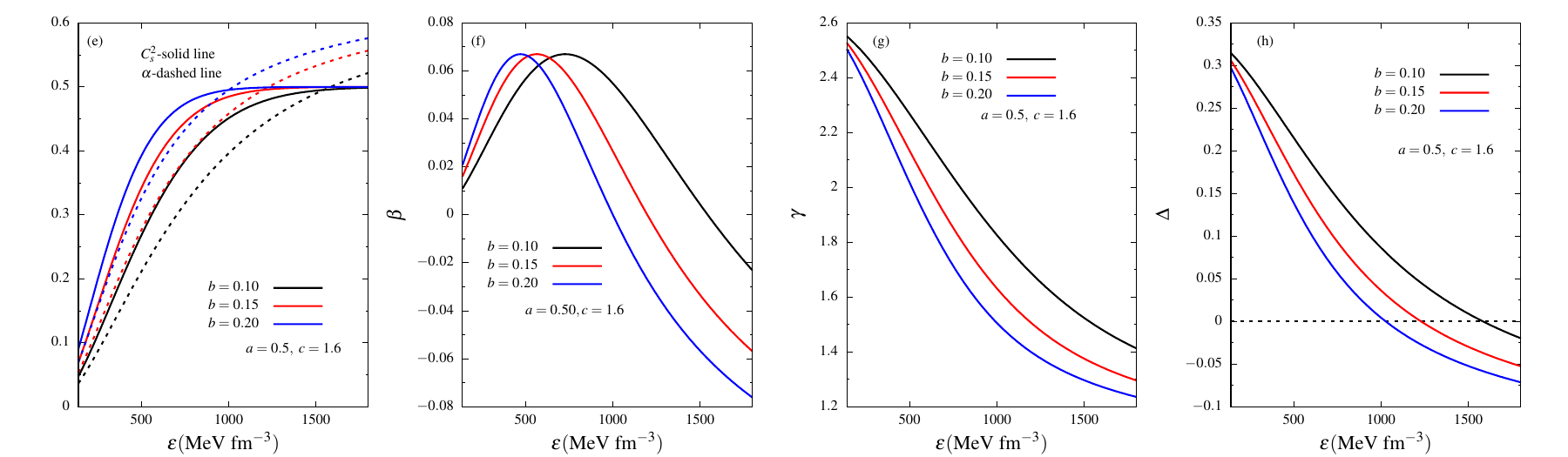}
    \includegraphics[width=0.90\textwidth]{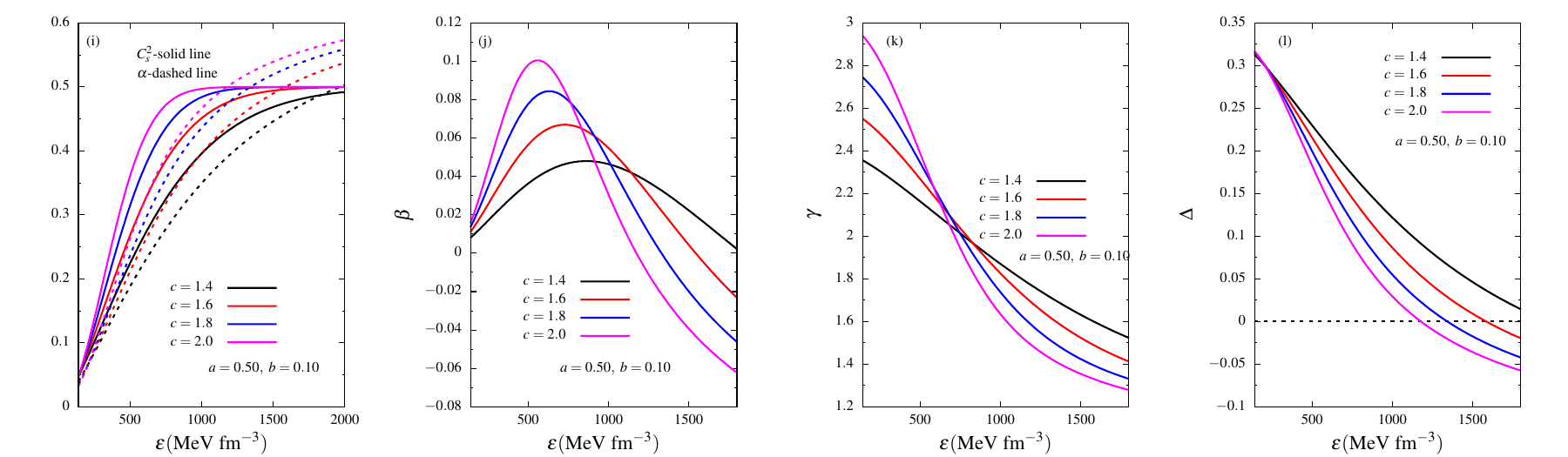}
    \caption{ 
    Left panel (a, e and i) shows the variation of $C_s^2$ and $\alpha$, left middle panel (b, f and j) displays the variation of $\beta$, right  middle panel (c, g and k) illustrates the variation of $\gamma$. and right panel (d, h and l ) shows the variation of the trace anomaly. The upper panel shows these variations for different values of a, the middle panel displays the variations for different values of b, and the lower panel presents the variations for different values of c.}   
    \label{fig:alpha_beta_var}
\end{figure*}  

\begin{figure*}[htp] 
    \centering
    \includegraphics[width=0.95\textwidth]{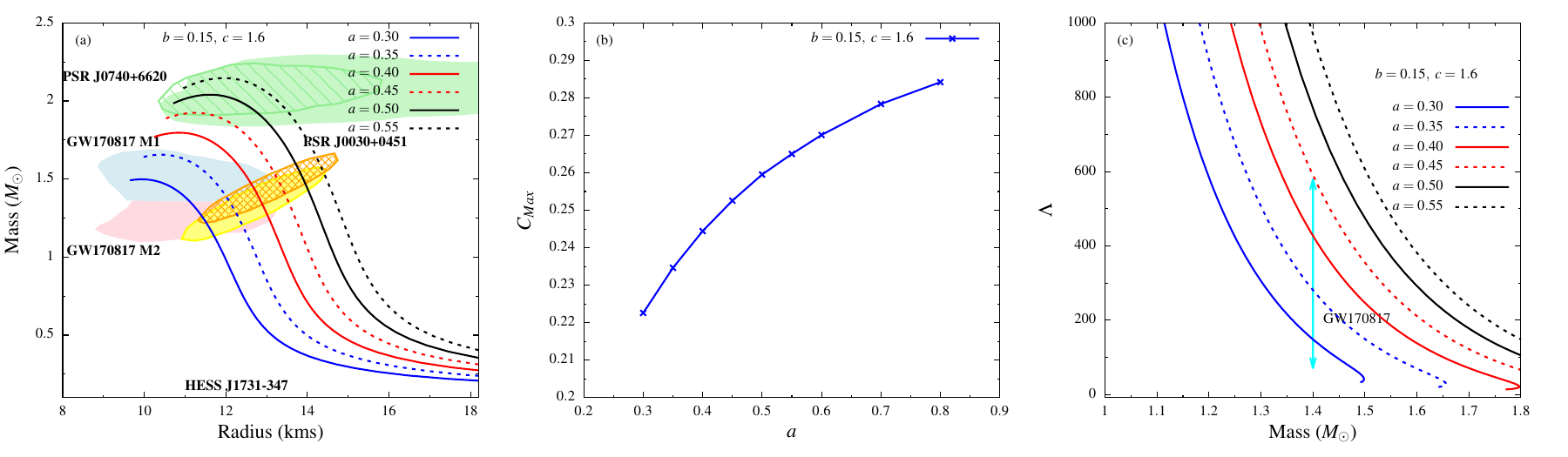}
    \includegraphics[width=0.95\textwidth]{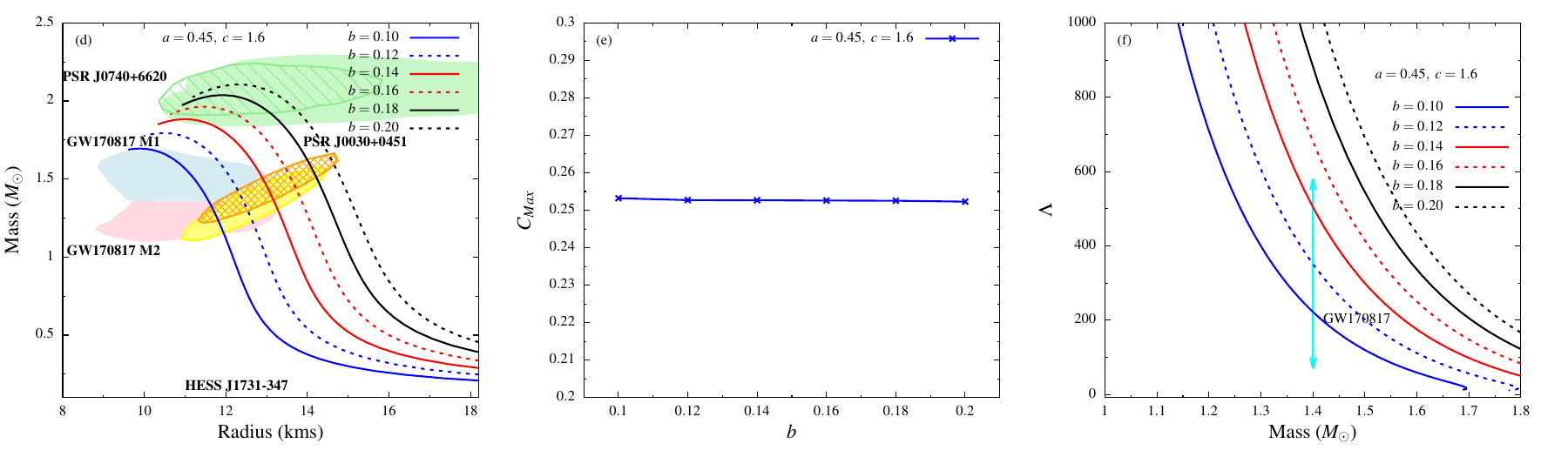}
    \includegraphics[width=0.95\textwidth]{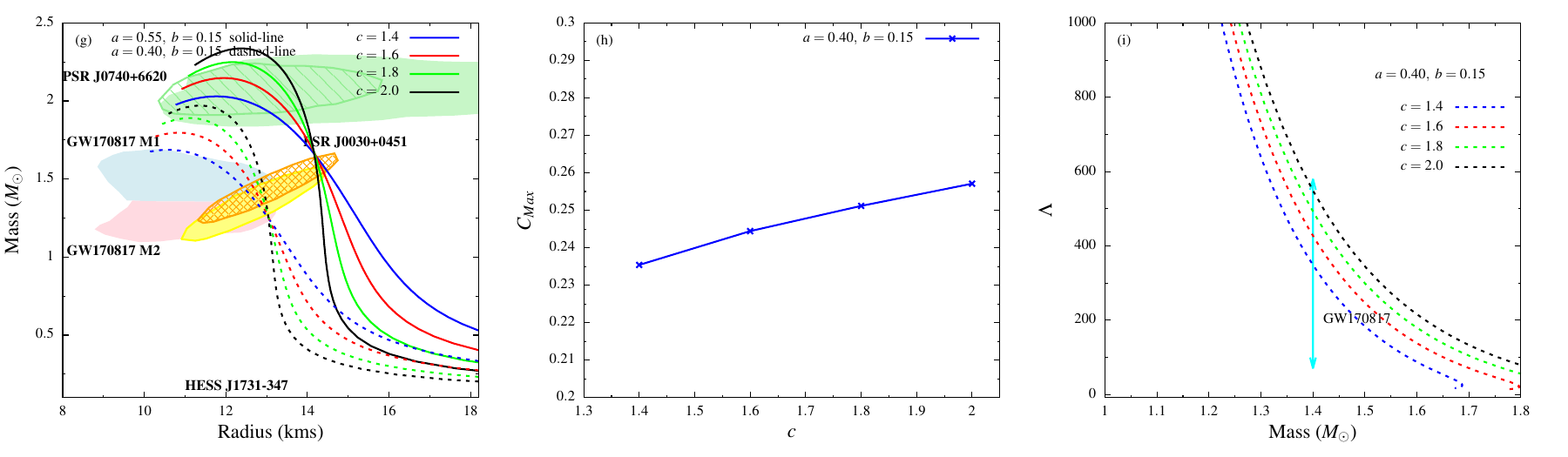}
    \caption{ Left panel (a, d and g) shows the  mass-radius relationship for the energy density-dependent speed of sound parametrization equation of state. Observational limits imposed from PSR J0740+6620\cite{Fonseca:2021wxt}, PSR J0030+0451 \cite{Riley:2019yda} indicated. The constraints on the $M-R$ plane prescribed from GW170817\cite{LIGOScientific:2018cki}  are also compared. The middle panel (b, e, and h) shows the variation of the compactness for the maximum mass and corresponding radius with the parameters a, b, and c respectively. (c) The right panel (c, f and i) shows the variation of tidal deformability with mass for different values of the above-mentioned parameters. The constraints on $70\le\Lambda_{1.4}\le 580$\cite{LIGOScientific:2018cki} are included. }   
    \label{fig:mr_lam}
\end{figure*}

%======================================
\section{Results} 
As has been mentioned earlier, in this work we introduce a parametrized density-dependent speed of sound in order to describe the physics of neutron star. In order to establish the merit of these parametrized equations of state, we have compared these with a few realistic hadron equations of state being calculated using the relativistic mean field theories. We have considered both the density-independent (NL3 \cite{nl3}, GM1 \cite{gm1}, IUFSU \cite{iufsu},FSU2R \cite{Tolos:2016hhl}) and density-dependent (DD2\cite{dd2}, DDME2 \cite{ddme2}) hadronic equation of state. We have tried to fit the realistic EOS with this parametrized one and have plotted both the original as well as the fitted ones in Fig.~\ref{fig:fit_cs2}. The fitted parameters are given in the Table.~\ref{tab:1}. The equations of state in the left figure are pretty close to the original while the speed of sound being plotted in the right panel reflects some difference between the original and the fitted ones for some of the EOS used. The difference is more prominent for the NL3 parametrization and least for that of the IUFSU one. We have used these real as well as the parametrized EOS (fitted with real ones) to calculate the maximum mass $M_{max}$ and the radius corresponding to the maximum mass $R_{max}$  and compared them in the Table.~\ref{tab:1}. It is seen that the maximum mass and the radius obtained are pretty close for the parametrized and the real one thus establishing the merit of the parametrization used. 

As we have already mentioned in the formalism section, using the thermodynamic relations, the speed of sound can be expressed as a sum of two terms, one proportional to the first derivative (slope, $\alpha$) of energy density and the 2nd being proportional to the 2nd derivative (curvature, $\beta$) of energy density. The first derivative reaches a maximum at some point which causes the curvature to vanish and it subsequently changes its sign from positive to negative.  The first term in the right-hand side of the Eq.~\ref{eq:cs2_decompose} exhibits a monotonically increasing pattern with density and when it subsequently equals and then surpasses the value of $c_s^2$, the 2nd term(curvature) changes its sign from positive to negative passing through the zero value. We have demonstrated this in Fig.~\ref{fig:iufsu_curv}, for a realistic equation of state (EOS) based on the RMF calculations with IUFSU parametrization \cite{iufsu}. This is a pure hadronic equation of state and the parameters $\alpha$ and $\beta$ behave as shown in Fig.~\ref{fig:iufsu_curv}(b). 
We have plotted $\alpha$, $\beta$, $\gamma$ and $c_s^2$ in the same plot in the right side of Fig. ~\ref{fig:iufsu_curv}(b) for the IUFSU parametrization. Some of the parameters have been appropriately scaled in order to fit in the same figure. It is observed that the value of density ($\rho$ = 0.7 $\text{fm}^{-3}$ )where the speed of sound equals the value of $\alpha$,  the parameter $\beta$ becomes zero and then subsequently changes its sign. 
%==========================================

In Fig.~\ref{fig:iufsu_curv}, we presented results only for the IUFSU EOS. For a more comprehensive analysis of the thermodynamic properties, we now compare all six selected RMF EOSs. As illustrated in Fig.~\ref{fig:trace_rmf}(a), the sign change in the second derivative  is sensitive to the EOS stiffness: stiffer EOSs exhibit maxima in their first derivatives at higher densities.
In Fig.~\ref{fig:trace_rmf}(e), we examine the trace anomaly, an important quantity associated with RMF models. Thermodynamic stability and causality impose the conditions $P>0$ and $P \leq \varepsilon$, respectively. In the case where $\varepsilon = 3P$, the quantity $\Delta$ vanishes ($\Delta = 0$). Consequently, $\Delta$ is constrained to the range $-2/3 \leq \Delta < 1/3$. The relationship between the trace anomaly and the stiffness of the equation of state (EOS) is well-reflected in different relativistic RMF models. We observe that for softer equations of state, such as IUFSU and FSU2R, $\Delta$ does not become negative, whereas for stiffer equations of state like NL3, GM1, DD2, and DDME2, the trace anomaly becomes negative at high densities. Our results for the trace anomaly show behavior which overlaps with the results reported in Ref.\cite{Annala:2023cwx}.
In Fig.~\ref{fig:trace_rmf}(f), we analyze the logarithmic rate of change of the trace anomaly $\Delta$ with respect to energy density. $\Delta^{\prime}$ is negative for all EOS considered, but its magnitude distinguishes between soft and stiff EOS: stiffer EOS exhibit more negative $\Delta^{\prime}$ values as compared to softer ones.
In Fig.~\ref{fig:trace_rmf}(g), we examine the combined quantity $d_c$, which spans a range of $[0, 0.45]$. Notably, softer EOS exhibit lower values of $d_c$ as compared to stiffer EOS. Our hadronic models predict the range of  polytropic index ($\gamma$)  from 1.15 to 3.5 (Fig.~\ref{fig:trace_rmf}(d)), which overlaps  with the results reported by Annala et al. \cite{Annala:2019puf}. To facilitate a better understanding of the polytropic index and the trace anomaly, Fig.~\ref{fig:3d_gamma_delta} presents 3D plots showing the variation of $\rho$, $C_s^2$, and $\gamma$ in the left panel, and $\rho$, $C_s^2$, and $\Delta$ in the right panel for the six RMF models. The relative stiffness of the different RMF models as well as the variation of the thermodynamic variables with density is clearly visible from the plots.
 It is observed from Fig.~\ref{fig:trace_rmf}(d,e)  that even if $\Delta$ becomes zero, $\gamma$ is not equal to  $1$ as $C_s^2$ differs from $1/3$  as  can be seen from  Fig.\ref{fig:fit_cs2}(b).

From the study of $\beta$ and $\Delta$, we have found that the sign change in $\beta$ is determined by the density at which $\alpha$ equals the speed of sound and by the trace anomaly when $P > 3\varepsilon$. For a softer equation of state, the sign change in $\beta$ occurs at a lower density as compared to  stiffer equation of state, while the opposite behavior is observed in the case of $\Delta$.

We have also calculated these parameters for our parametrized equations of state as given by Eq.~\eqref{eq:cs2params}.  The nature of variation remains the same irrespective of the chosen parameters as seen from Fig.~\ref{fig:alpha_beta_var}.
Here also we have examined by varying one parameter at a time, keeping the other two fixed. In the left panel (Fig.~\ref{fig:alpha_beta_var}(a),(e),(i)), we have plotted the speed of sound and $\alpha$ and it is observed that $\alpha$ shows a monotonic increase and becomes equal to the speed of sound at a certain value of energy density which depends of course on the parameter values. The energy density where this crossing happens increases as the value of parameter $a$ is being increased.  In the Fig.~\ref{fig:alpha_beta_var}(e), we vary $b$ keeping the other two fixed and it is observed that the crossing of $c_s^2$ with $\alpha$ happens at lower energy densities as 
$b$ is increased. In the Fig.~\ref{fig:alpha_beta_var}(i), we have varied the parameter $c$  and the results are similar to the case of varying parameter $b$. 
The left middle column (Fig.~\ref{fig:alpha_beta_var}(b),(f),(j))shows the variation of $\beta$  which exhibits a Gaussian pattern with energy. $\beta$ (curvature term) changes its sign in all cases irrespective of the saturation value of the speed of sound $a$. This is seen from Fig.~\ref{fig:alpha_beta_var}(b), where we change the parameters $a$ keeping two other parameters fixed.
 Increasing the parameter $a$ and keeping $b$ and $c$ fixed,  the peak shifts its position to higher energy and there is a simultaneous increase in height too. In the Fig.~\ref{fig:alpha_beta_var}(f),  it is observed that the peak in  $\beta$ shifts to the right as one decreases the parameter $b$, the height remaining more or less constant irrespective of the change in parameter $b$. With the increase in parameter $c$, the peak position in $\beta$ shifts to the left towards lower energy densities, the height getting increased, but the width of the curves decreased. From the lower figure (Fig.~\ref{fig:alpha_beta_var}(j)), one can approximately conclude that the area under the curves changes little as the parameter $c$ is changed.

 We study the variation in the polytropic index  ($\gamma$) with energy density in right middle panel (Fig.~\ref{fig:alpha_beta_var}(c),(g),(k)). From the Fig.~\ref{fig:alpha_beta_var}(c),  it is observed that $\gamma$ decreases with energy density, the interesting part being, this change being completely independent of the variation in the parameter $a$.  This is expected from the definition of $\gamma$ in the Eq.~\eqref{eq:poly_index}; after substituting, one can see that it becomes independent of the parameter $a$ which represents the saturation value of the speed of sound.  In the Fig.~\ref{fig:alpha_beta_var}(g), the change is being displayed for different values of the parameter $b$, the other two being kept constant. In this case though the pattern of decrease in $\gamma$ remains same,  the values depend on the parameter $b$. In the Fig.~\ref{fig:alpha_beta_var}(k), we study the variation in $\gamma$ varying the parameter $c$ and it is observed that gamma decreases, the pattern being dependent on the value of $c$, the lowest value of $c$ resulting in the least change in $\gamma$.

We study the variation in the trace anomaly with energy density for the parametrized $C_s^2$ EOS shown in the  Fig.~\ref{fig:alpha_beta_var}(d),(h),(l)).  From the Fig.~\ref{fig:alpha_beta_var}(d), we have found that $\Delta$ changes sign for the stiffer equation of state  at lower densities as compared to the softer ones with lesser values of  the parameter $a$  which controls the stiffness of the EOS. $\Delta$ changes sign for all our chosen values of the  the parameter $b$ within the selected range of energy density. For the parameter $c$ the behavior of $\Delta$  is somewhat  similar to that of the parameter $a$.

Next,  we use this parametrized equations of state to calculate the structural properties of the compact stars. First, we would like to examine the effect of each of these parameters on the mass-radius (M-R) diagram by varying one parameter at a time, keeping the other two fixed. In Fig.~\ref{fig:mr_lam}(a)(upper panel), the effect of variation of $a$ is studied keeping both $b$ and $c$  fixed.
  An increase in the value of this parameter $a$ results in increase of both maximum mass $M_{max}$ as well as radius corresponding to $M_{max}$.  In the next figure, we fix the values of $a$  and $c$ and vary $b$. Both maximum mass $M_{max}$ and the corresponding radius increase with the increase in the values of $b$ as in the case of $a$ though the magnitude of change is different. Finally, we vary the parameter $c$ keeping $a$ and $b$ fixed. In this case too, $M_{max}$ and the corresponding radius increase as $c$ is increased but here the M- R plots cross each other at a point unlike the two previous cases. This is a special point that has been observed earlier in the case of hybrid stars. Here it is observed that varying the parameter $c$, one obtains the solutions in M-R which always cross each other at a special point irrespective of the values of the other two parameters. The M-R results satisfy the constraints from PSR J0030+0451 for  almost all the values of parameter $a$, $b$ and $c$ as seen from the figures. The constraints from GW170817 are satisfied for the lower values of the parameters $a$ and $b$.  The data from PSR J0740+6620 are satisfied by higher values of the parameters. 

The compactness parameter $C_{max}=\frac{M_{max}}{R_{max}}$  increases as $a$ is increased, as is seen from Fig.~\ref{fig:mr_lam}(b).  The compactness parameter however is observed to be almost independent of $b$ as seen from the Fig.~ \ref{fig:mr_lam}(e). The compactness parameter increases with $c$ as is observed from the Fig.~\ref{fig:mr_lam}(h)though the increase is less as compared to that of $a$.

Tidal deformability (TD) is an important observable which constraints the equations of state to a great extent and hence we have calculated the same in this work with our ensemble of EOS in order to restrict the range of parameter values. TD puts a cut on the higher values of the parameter $a$ (Fig.~\ref{fig:mr_lam}(c) and restricts it upto a maximum value which depends on the other two parameters $b$ and $c$. This is expected as we are already aware that steep equations of state like NL3 fails to satisfy the tidal deformability constraint. When the parameters $a$ and $c$ are being fixed and the parameter $b$ is varied, it is observed that TD restricts the value of $b$ (Fig.~\ref{fig:mr_lam}(f)) to a certain maximum which again depends on $a$ and $c$. Similar feature is observed from Fig.~\ref{fig:mr_lam}(i) when the parameter $c$ is being varied.

\section{Conclusions} 
In this work, we have introduced a parametrized density-dependent speed of sound in order to describe the neutron star equation of state in a different approach. The energy dependence of  $C_s^2$  is defined by three parameters $a$, $b$, and $c$, each displaying some special feature with respect to different properties of the neutron star. The polytropic index $\gamma$ is observed to be independent of the parameter $a$ which is connected to the saturation value of $C_s^2$. The compactness of the neutron stars corresponding to the maximum mass
is found to be independent of the parameter $b$. The variation in parameter $c$ generates EOS which yields special points in the Mass-Radius plot of the neutron stars, a feature which has been earlier observed mostly in the case of hybrid stars. 

 One motivation of our work is analysis of the behavior of the speed of sound and other thermodynamic quantities.  We have demonstrated  that for a real hadronic equation of state 
 the first derivative of energy density per baryon density wrt density attains a maximum at a certain density, where the second derivative of energy density per baryon density vanishes. This clearly establishes that the sign change in  $\beta$ is neither related to first-order quark hadron phase transition nor the matter at high density reaching the conformal limit. We have also studied the behavior of the trace anomaly. The sign change of $\beta$ and $\Delta$ also depends on the stiffness of the hadronic equation of state. For a stiffer equation of state, the sign change in $\beta$ may occur at a higher density, whereas the sign change in $\Delta$ can take place at an earlier transition density. We have also studied another important quantity, the polytropic index. We found that the values of $\gamma$ do not reach the conformal value of 1, while $\Delta$ changes sign. This implies that the quantity $\gamma$ is more important in the context of  identification of the conformal limit. We would like to point out explicitly that not only do our results align with existing literature, but they do so using purely hadronic models. This is a key novelty of our work and as it demonstrates that certain thermodynamic behaviors often associated with deconfined quark matter or phase transitions---can arise without invoking an explicit quark-hadron phase transition.

 We have also study the thermodynamics for the parametrized $C_s^2$ model. This sign change in $\beta$ and $\Delta$ happens at different energy densities depending on the values of $a$, $b$, and $c$  when the speed of sound equals $\alpha$  irrespective of any other constraint on the composition of the matter. We have studied in detail the behavior of thermodynamic quantities like $\alpha$, $\beta$, $\Delta$ and the polytropic index  $\gamma$ with the energy density for the parameters $a$,$b$, and $c$. $\alpha$ has a monotonically increasing behavior  while $\beta$ shows a Gaussian type variation with energy density.  We have also shown that the dimensionless tidal deformability of 1.4 $M_{\odot}$ can be used to constrain the range of the parameters used in defining the speed of sound.

\bibliographystyle{elsarticle-num}
\bibliography{bibliography}
\end{document}